\input{aipcheck}

\documentclass[
    ,final            % use final for the camera ready runs
  ]
  {aipproc}

\layoutstyle{6x9}

\begin{document}

\title{Determination of electroweak parameters at HERA with the H1 experiment}


\author{Benjamin Portheault}{
  address=Laboratoire de l'Acc\'el\'erateur Lin\'eaire\\ IN2P3-CNRS et Universit\'e de Paris Sud\\ F-91898 Orsay Cedex\\ E-mail: portheau@lal.in2p3.fr}

\begin{abstract}
Using deep inelastic $e^{\pm}p$ charged and neutral current scattering cross sections previously published, a combined electroweak and QCD analysis is performed to determine electroweak parameters accounting for their correlation with parton distributions. The data used have been collected by the H1 experiment in 1994-2000 and correspond to an integrated luminosity of $117.2$ pb$^{-1}$. The $W$ boson mass is determined via the propagator and in the on-mass-shell scheme. A first measurement at HERA is made of the light quark weak couplings to the $Z^0$ boson.
\end{abstract}

\maketitle

%%%%%%%%%%%%%%%%%%%%%%%%%%%%%%%%%%%%%%%%%%%%
%% MAINMATTER
%%%%%%%%%%%%%%%%%%%%%%%%%%%%%%%%%%%%%%%%%%%%

\section{THE HERA INCLUSIVE DIS DATA}

The H1 experiment has collected $117.2$ pb$^{-1}$ of data during the first phase of HERA operation (HERA I) with  unpolarised $e^{\pm}$ beams. The inclusive deep inelastic scattering (DIS) cross section \cite{H1NcCcEp,H1lowq9697,H1NcCcEm,H1NcCcQcdFit} have been measured in neutral current (NC) and charged current (CC). The precise low $x, Q^2$ data became the cornerstone of any QCD analysis (the so called QCD fits) aiming at extracting the parton distribution functions (PDFs). The addition of CC data allowed the flavor separation of PDFs and thus a QCD analysis based solely on H1 data has been performed \cite{H1NcCcQcdFit}, with the advantage of a consistent treatment of correlated systematic errors.
%, a crucial issue when considering the determination of meaningful errors coming from the QCD analysis. While the QCD aspect of the cross section is now well determined, little effort has been devoted to the determination of electroweak (EW) parameters entering in the DIS cross sections using the full data set. 

\section{A COMBINED ELECTROWEAK-QCD ANALYSIS}

The combined EW-QCD analysis aims at determining EW and PDFs parameters in the same fit to NC and CC data. This removes a bias due to the fact that PDFs parameters are obtained with assumptions on EW parameters, so the use of fixed PDFs parameters to fit the EW ones is not a consistent procedure. The combined fitting allows to consistently take into account the uncertainty coming from the proton's structure. 

\subsection{THE H1PDF2000 QCD ANALYSIS}
The basis for the combined EW-QCD analysis is the H1 QCD analysis of HERA I data, the so-called H1PDF2000 \cite{H1NcCcQcdFit,qcdfit}. In this analysis five combinations are parameterized and fitted, the gluon $g$, the up-type quarks $U=u+c$, the down-type quarks $D=d+s$ and the anti-quarks distribution $\overline{U},\overline{D}$. The parameterization space is explored with a systematic procedure until a $\chi^2$ saturation is reached. The resulting paramterization has a total of 10 free parameters.

\subsection{DESCRIPTION OF THE ELECTROWEAK SCHEMES}

The choice of a renormalization scheme defines the set of theory parameters to be used as inputs for the cross sections calculations. For the CC and NC cross sections, two main schemes are used\,: the On Mass Shell scheme (OMS) which uses the weak boson masses as fundamental parameters, and the Modified On Mass Shell scheme (MOMS) in which the $W$ boson mass is replaced by the Fermi constant $G_F$. Since the Fermi constant includes by definition the radiative corrections due to the $W$ self energy, the CC cross section reads 
\begin{equation}
\frac{\mathrm{d}^2\sigma_{CC}^{e^\pm p}}{\mathrm{d}x\mathrm{d}Q^2}=\frac{G_F^{2}}{4\pi x}\left[\frac{M_{W}^{2}}{Q^2+M_{W}^{2}} \right]^2 \Phi^\pm_{\mathrm{PDFs}}(x,Q^2)
\label{CCmoms}
\end{equation}
in which $\Phi^\pm_{\mathrm{PDFs}}$ is the structure function part and $M_W$ has to be computed from $G$ and the other parameters. To use of the OMS scheme, one has to replace $G$ by its expressions as a function of masses and one needs to compute the relatively large ($\sim 3\%$) radiative correction  which is a function of all the Standard Model parameters $\Delta r (\alpha,M_Z,M_W,m_t,M_H)$. In particular $\Delta r$ has a quadratic dependency upon the top quark mass $m_t$ and a logarithmic dependency upon the Higgs mass $M_H$. The CC cross section reads
\begin{equation}
\frac{\mathrm{d}^2\sigma_{CC}^{e^\pm p}}{\mathrm{d}x\mathrm{d}Q^2}=\frac{\pi\alpha^{2}}{4M_W^2\left(1-\frac{M_W^2}{M_Z^2}\right)^2}\frac{1}{(1-\Delta r )^2}\left[\frac{M_{W}^{2}}{Q^2+M_{W}^{2}} \right]^2 \Phi^\pm_{\mathrm{PDFs}}(x,Q^2).
\label{CComs}
\end{equation}
The $W$ mass entering in Eq. \eqref{CCmoms} is called the \emph{propagator mass} as it enters only in the $Q^2$ dependency of the cross section. In the OMS scheme, the $W$ mass enters also in the normalization of the cross section, yielding an increased sensitivity to this parameter.

\section{RESULTS FOR THE PROPAGATOR MASS} 
\begin{figure}
  \includegraphics[height=.24\textheight]{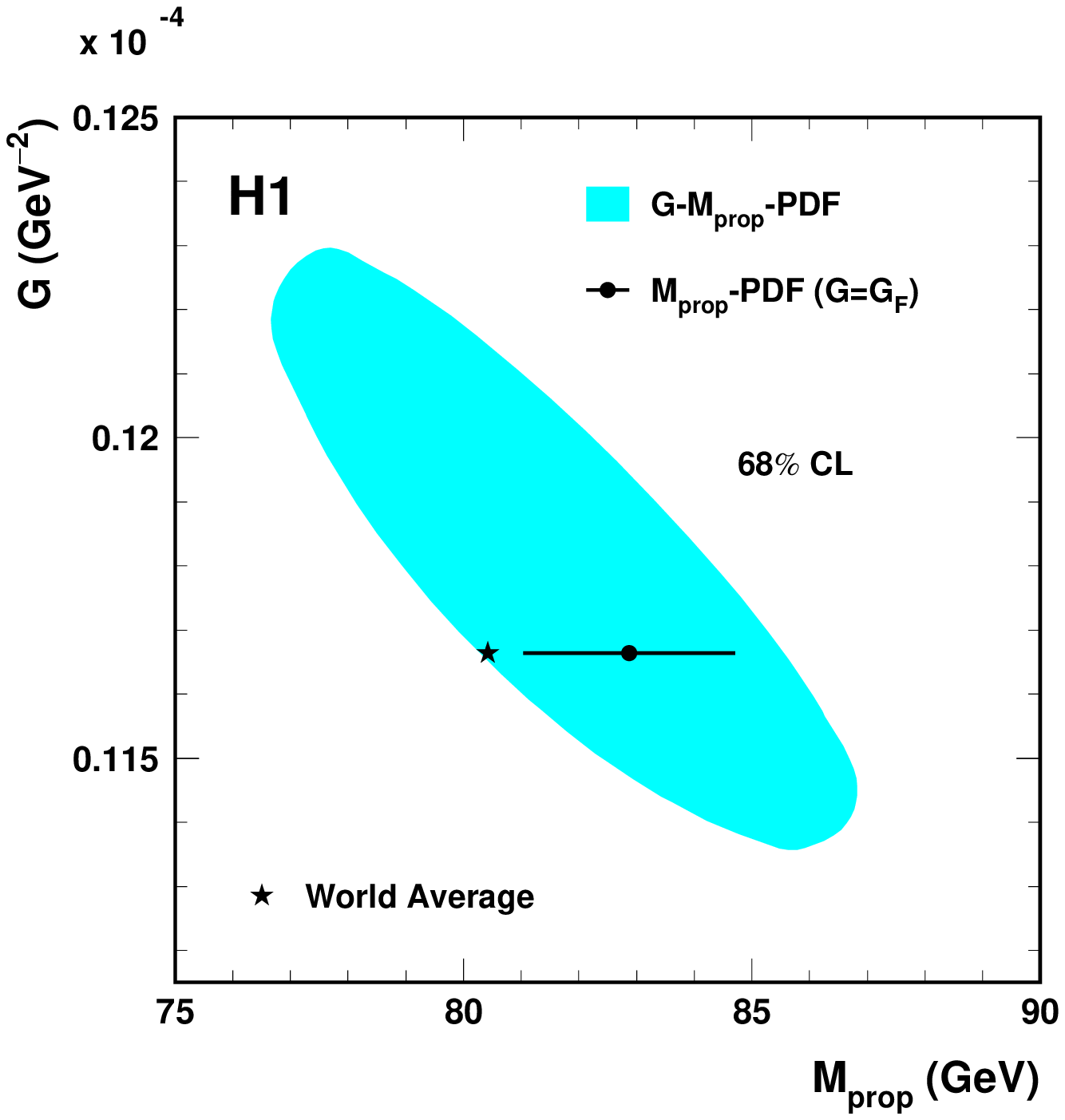}
 \hspace{5mm}
   \includegraphics[height=.22\textheight]{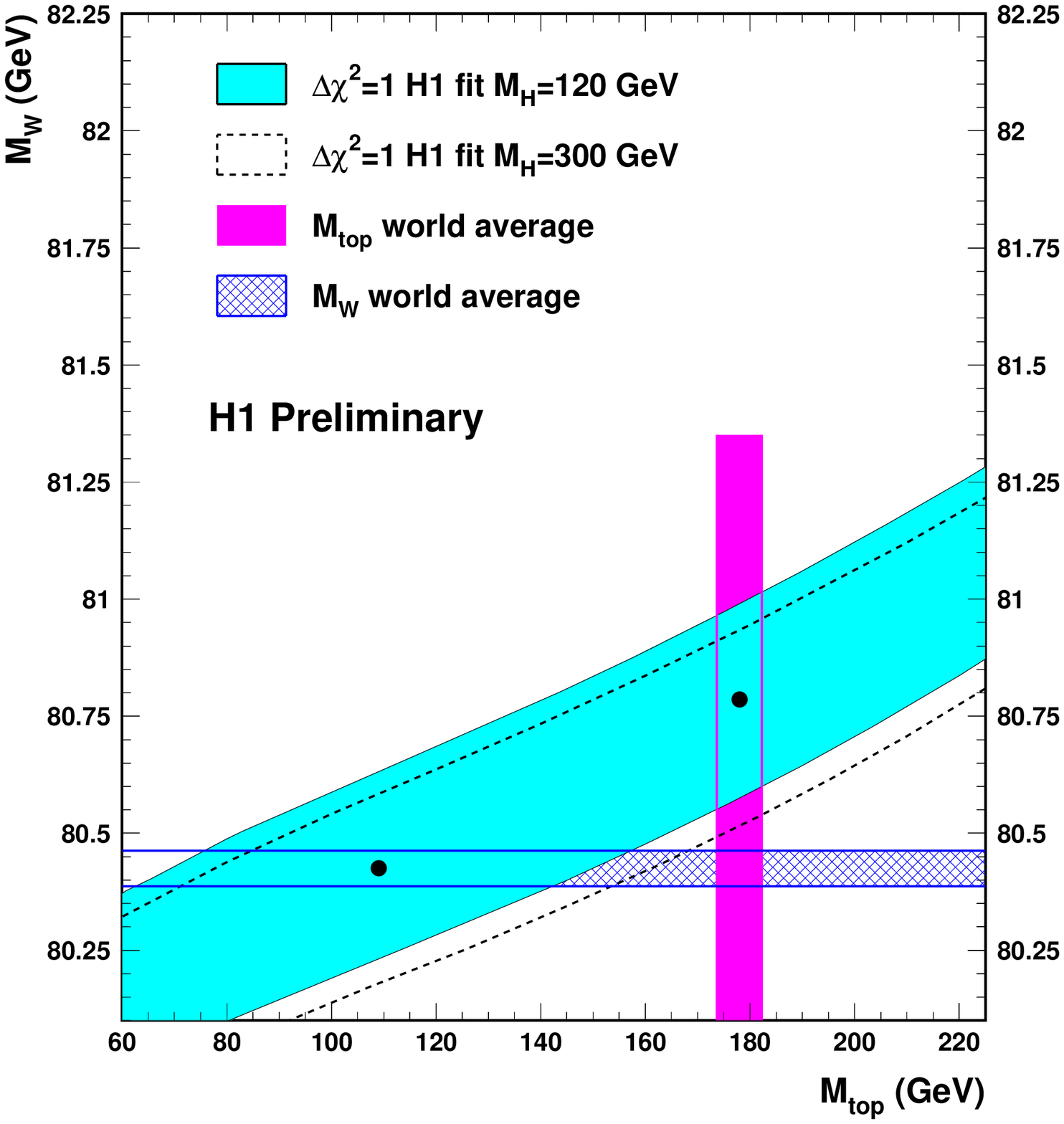}
  \caption{Left figure: result of the fit to $G$  and $M_{prop}$ at $68$\% CL. Fixing $G$ to $G_F$, the fit results in a measurement of the propagator mass $M_{prop}$ shown as the dot with the horizontal error bar. Right figure: result of the $m_t,M_W$ and PDFs fit in the $m_t,M_W$ plane for $M_H=120$ GeV (shaded area) and for $M_H=300$ GeV (dashed area). The vertical shaded band and the horizontal hatched band represent respectively the world average values of the top and $W$ masses.\label{fig:gmprop}}
%  \caption{Picture to fixed height}
\end{figure}

In a first analysis, we consider the propagator mass $M_{prop}$ and the normalization constant $G$ in Eq. \eqref{CCmoms} as independent. The result of a 12 parameter fit $G$-$M_{prop}$-PDFs is shown on Fig. \ref{fig:gmprop}.
Then the parameter $G$ is fixed to the well measured Fermi constant $G_F$ and the propagator mass is fitted together with the PDFs. The result is 
\begin{equation} 
M_W=82.87\pm 1.83(\mathrm{exp}) {}^{+0.30}_{-0.16}(\mathrm{mod})\ \mathrm{GeV}
\end{equation}
where the experimental uncertainty accounts for all statistical and systematic errors. The second error is due to model uncertainties, accounting for variations on the parameters entering in the QCD fit modelisation\cite{H1NcCcQcdFit}. Using fixed PDFs from H1PDF2000  would introduce a systematic bias of $0.5$ GeV on $M_{prop}$. This is the first coherent determination of the propagator mass in $ep$ collisions.

\section{OMS ANALYSIS}
In the OMS scheme one needs to compute the radiative correction to the $W$ self energy. This is done using EPRC program \cite{eprc}, and terms of order $\mathcal{O}(\alpha)$, $\mathcal{O}(\alpha \alpha_s)$ and leading  $\mathcal{O}(\alpha^2)$ are included. First, a 12 parameter fit of $M_W,m_t$ and PDFs is performed, and the result in the $M_W,m_t$ plane is shown in Fig. \ref{fig:gmprop}.
%\begin{figure}
%  \includegraphics[height=.3\textheight]{mw_mtop_dis05.eps}
%  \caption{Result of the $m_t,M_W$ and PDFs fit in the $m_t,M_W$ plane for $M_H=120$ GeV (shaded area) and for $M_H=300$ GeV (dashed area). The vertical shaded band and the horizontal hatched band represent respectively the top and $W$ masses world average. \label{fig:mwmtop}}
%  \caption{Picture to fixed height}
%\end{figure}
The choice of the Higgs mass shifts the allowed region. Since the top mass has been directly measured at the TeVatron, one can use its value to constrain the $W$ mass. The result for $m_t=178$ GeV, $M_H=120$ GeV is 
\begin{equation}
M_W=80.786\pm 0.207(\mathrm{exp}) {}^{+0.048}_{-0.029}(\mathrm{mod}) \pm 0.025(\mathrm{top}) \pm 0.033(\mathrm{th})\ \mathrm{GeV}.
\end{equation}
In addition to the experimental and model uncertainties, uncertainties from the top mass and theoretical uncertainty on $\Delta r$ calculation have been taken into account\cite{thesis}. Using a Higgs mass of $300$ GeV instead of $120$ GeV changes $M_W$ by $-0.084$ GeV. One needs to emphasize that this is not a measurement of the $W$ mass but a model dependent determination of a Standard Model parameter, because the validity of Standard Model is assumed is to compute the radiative correction. Given that the result is $1.7 \sigma$ from the world average value of the $W$ mass, the H1 data are consistent with the Standard Model of strong and electroweak interactions. 
This result can be translated into a determination of $\sin^2\theta_W=1-M_W^2/M_Z^2$ in the OMS scheme, using the $Z^0$ mass world average. The result is $\sin^2 \theta_W=0.2151\pm 0.0040 (\mathrm{exp}){}^{+0.0019}_{-0.0011}(\mathrm{th})$. 
It is also possible to fit other parameters via the radiative correction, such as the top or the Higgs mass. The result is $m_t=108\pm44$ GeV, and $\log_{10} M_H=3.9\pm 2.2 (\mathrm{exp})$. One thus obtains a top mass consistent with direct measurement and a Higgs mass larger than $50$ GeV at $68 \%$ CL via the radiative corrections. This is the first time such a determination is obtained at HERA. 

\section{QUARKS COUPLINGS TO THE $Z^0$}
\begin{figure}
  \includegraphics[height=.25\textheight]{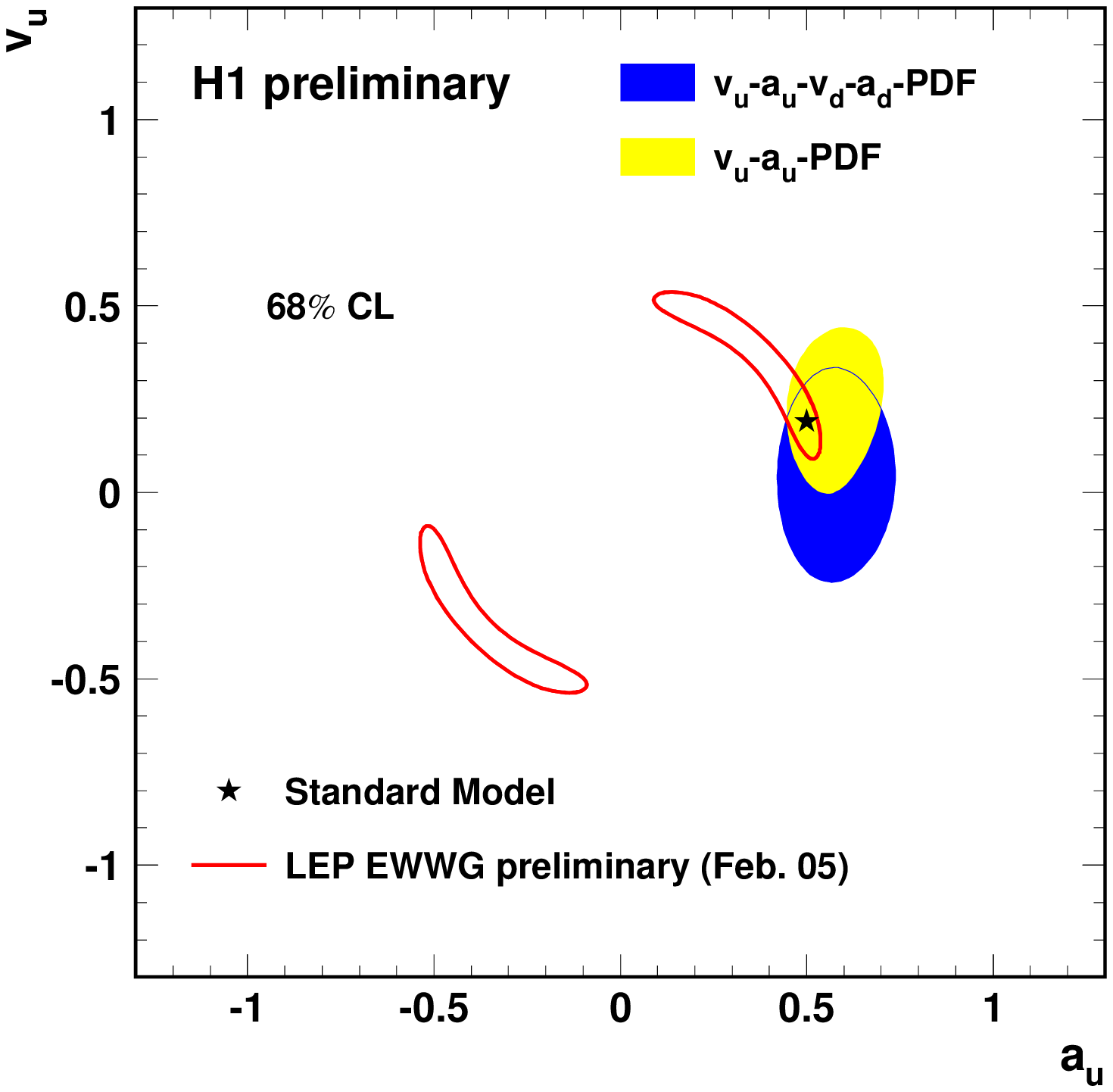}
  \includegraphics[height=.25\textheight]{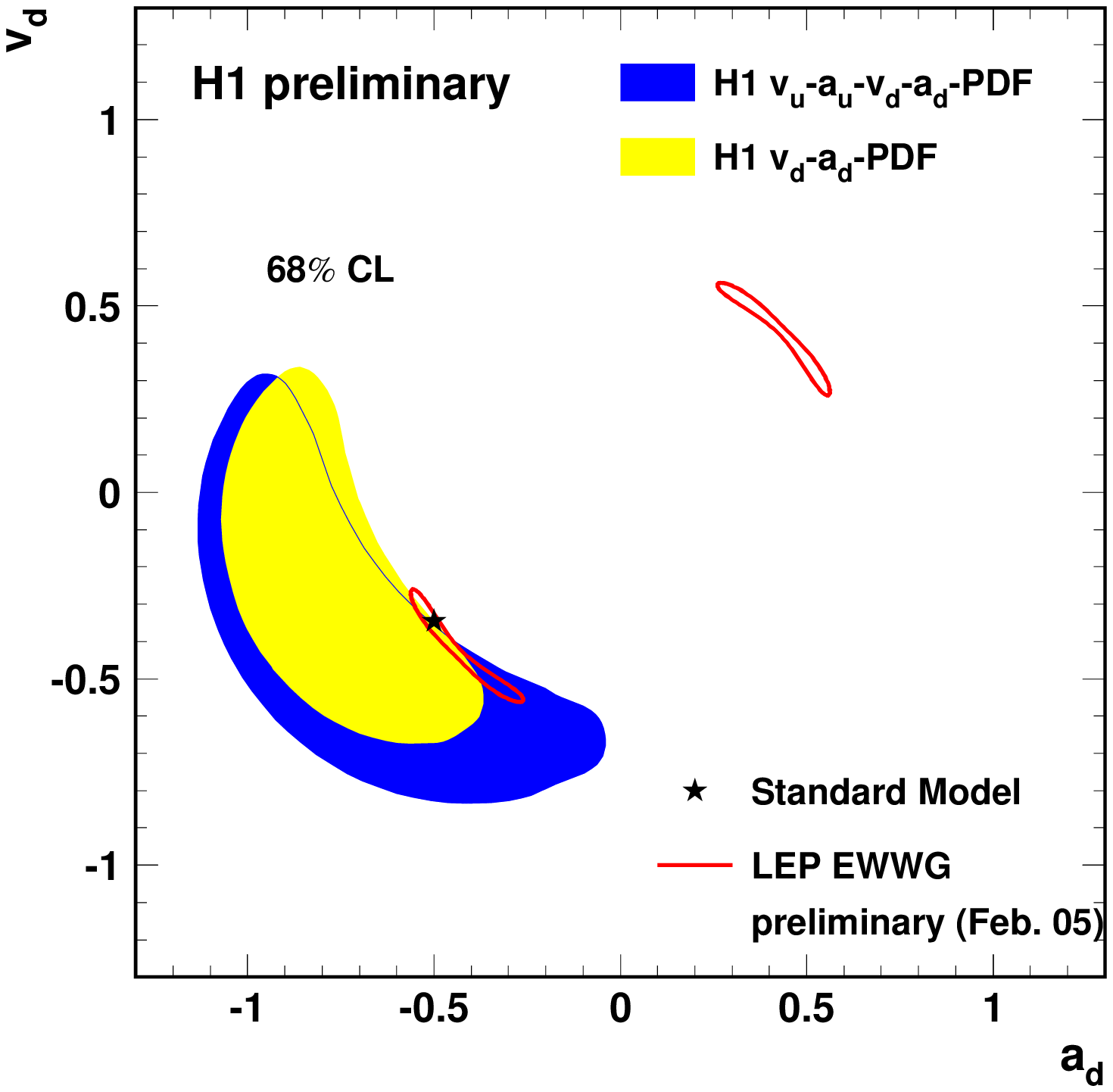}
  \caption{Results at $68$\% CL on the weak neutral current couplings of $U$ (left plot) and $D$ (right plot) quarks to the $Z^0$ boson determined in this analysis. The dark-shaded contours correspond to results of a simultaneous fit of all four couplings. The light-shaded contours correspond to results of fits where either $d$ or $u$ quark couplings are fixed to their SM values. The open contours are from the combined LEP data\cite{lepcouplings}.\label{fig:couplings}}
\end{figure}

The axial $a_q$ and vector $v_q$ couplings of quarks to the $Z$ have been measured precisely for heavy quarks in $e^+ e^-$ collisions, but are less well measured for light quarks. In DIS, combinations weighted by partons densities of the four couplings $a_U, a_D, v_U, v_D$ appear in the $\gamma-Z^0$ and pure $Z^0$ terms of $F_2$ and $xF_3$. It is possible to extract unambiguously these couplings together with the parton densities in a combined fit. The results are shown in Fig. \ref{fig:couplings}. 
This is the first HERA result on this topic, and a significant future improvement is expected with the HERA II polarised data.

%To conclude, the use of a new combined EW-QCD analysis allows to extract electroweak parameters with the H1 HERA I data. In particular the $W$ mass is measured via the propagator mass and the $W$ mass parameter is determined in an OMS analysis. The first measurement of light quarks couplings to the $Z$ boson in $ep$ collisions is performed.
%All these results will be improved with the use of the HERA II data.

%%%%%%%%%%%%%%%%%%%%%%%%%%%%%%%%%%%%%%%%%%%%%%%%
%% BACKMATTER
%%%%%%%%%%%%%%%%%%%%%%%%%%%%%%%%%%%%%%%%%%%%%%%%

%\begin{theacknowledgments}
%  bla bla 
%\end{theacknowledgments}



\bibliographystyle{aipproc}   % if natbib is available

%\bibliography{sample}

%  \typeout{******************************************}
%  \typeout{** Please run "bibtex \jobname" to optain}
%  \typeout{** the bibliography and then re-run LaTeX}
%  \typeout{** twice to fix the references!}
%  \typeout{******************************************}

\end{document}